# Relevance of IsoDAR and DAEδALUS to Medical Radioisotope Production

Jose R. Alonso
*Massachusetts Institute of Technology*

**Abstract**

The very-high current cyclotrons being designed for the IsoDAR and DAEdALUS experiments are of value to fields outside of neutrino physics. In particular, the medical isotopes industry can benefit from these cyclotron developments to produce a new generation of machines with capabilities far in excess of today's technology. This paper provides a tutorial on the field of medical isotopes: from properties of isotopes desired for clinical applications, to production considerations and available technology, concluding with discussion of the impact of the new cyclotrons on the field.

## Introduction:

Medical uses of radioisotopes fall into two categories: diagnostic and therapeutic. The isotope is introduced into the patient in a form that will be selectively absorbed by the organ or tissue that is targeted; either in elemental form (such as iodine isotopes targeting the thyroid gland), or attached to a pharmaceutical which carries the radioactive tracer to its targeted site (e.g. FDG – fluoro-deoxyglucose; the deoxyglucose is a metabolite – sugar – that is absorbed from the blood in tissue undergoing high metabolic activity, such as active areas of the brain, or rapidly-growing tumors; the fluorine tracer goes along for the ride but serves as a marker to the location of the absorbed material).

Many different isotopes are now in clinical use, produced by a variety of techniques. However, many highly-desirable isotopes are not being used today because of production issues, or technological limitations. Many studies have been undertaken to address the complex technological and political issues with enhancing the availability of desirable isotopes [1,2,3,4] but progress is slow. As will be shown, the cyclotron-development efforts connected with the DAEδALUS and IsoDAR projects can play a very important role in addressing this problem.

But first, we will provide background relating to the medical uses of radio-isotopes, and factors relevant to these applications.

## Isotope Properties and Medical Procedures

Properties important for medical use are: decay modality (alpha, beta, positron, gamma, x-ray, neutron, fission); half-life; and means of production.

**Decay Modality**:

Once the isotope has been delivered to its target site, its mission may be either to destroy tissue in close proximity to it (therapeutic) or to provide information to the clinician of the location, size and shape of the tissue containing the activity (diagnostic).

*Therapeutic application:*

To deliver a therapeutic dose, the radiation damage should be localized to the immediate region containing the activity. The most effective isotope will undergo alpha decay, or beta decay, and will have very little gamma component. Low-energy electrons emitted through the Auger process are also effective at depositing energy close to the decaying nucleus.

*Diagnostic application:*

On the other hand, for diagnostic imaging the isotope should deliver as little dose as possible to the tissue, but should decay with gamma or x-rays that are not significantly attenuated in the body and are efficiently detected by segmented radiation counters. Segmentation (or another form of position-sensitivity) is important to localize the source of the radiation. Photon energies of 100 to 300 keV are most useful; little is absorbed in tissue, and detection efficiency is high in germanium, sodium-iodide, bismuth-germanate or other gamma-ray detectors.

<u>PET (Positron-emission tomography)</u>

Beta-plus (positron) emission from a nucleus provides excellent imaging opportunities. The positron slows down by collisions with electrons in the vicinity of the decay, and eventually stops after traveling an average of a millimeter or two from its source. At the stopping point it annihilates with an electron producing two 511 keV gamma rays that are (almost) exactly 180° aligned. Detecting the coincident gammas defines a line through the site of the decaying nucleus. Observing many decays allows reconstruction of the locus of activity in the patient. Annihilation radiation energy (511 keV) is a little high for efficient detection in small (highly-segmented) detector sections, but this is not a show-stopper for PET imaging. Just makes it harder.

<u>SPECT (Single-photon emission computerized tomography)</u>

In the absence of a coincident pair, determining the direction of the detected photon requires a special "camera" with a honeycomb collimator in front of the detectors. While transmission through the collimator is substantially reduced from absorption of off-axis photons, good images can be obtained with sufficient source-strength and statistics.

**Half-lives:**

Ideally, the radiation source would be turned on during the time the procedure is taking place, and then is turned off. In some therapeutic applications, known as brachytherapy, radioactive seeds are implanted in the tumor site, left until the prescribed dose is delivered, and then might be removed. This is not generally feasible for sites where the isotope is delivered through the bloodstream, or where the activity cannot be removed following the procedure. In this case, the half-life of the isotope should be short enough to minimize the dose delivered to the patient following the procedure.

On the other hand, the isotope begins decaying as soon as it is produced, so the time to prepare and administer the isotope should be short compared to its half-life. A typical compromise is that half-life should be no longer than a few hours for an isotope used in a diagnostic procedure.

Best is if the isotope can be produced very close to the final use-site. Two common methods of accomplishing this are: small cyclotrons, and generators.

Small cyclotrons (delivering a few tens of microamps of 12-15 MeV protons) in the hospital or clinic can produce light positron-emitters – e.g. $^{18}$F (110 minute half-life), $^{11}$C (20 minute half-life) – and auto-chemistry systems deliver labeled compounds such as FDG within a few minutes that are ready for injection into the patient.

A "generator" is a radioactive source (parent), with reasonably-long half-life (greater than a few days), with a daughter product that is used in the clinical procedure. For instance, the well-known $^{99}$Mo/$^{99m}$Tc generator is one of the most wide-spread nuclear-medicine tools in use today (over 50,000 procedures performed daily in US). $^{99}$Mo has a 66 hour half-life (2.5 days), the $^{99m}$Tc daughter lives 6 hours. A clinic will receive a shipment of $^{99}$Mo – a very common fission fragment from uranium reactors with HEU (highly-enriched uranium) cores – perhaps once a week; and a dose of $^{99m}$Tc is "milked" as needed just prior to a planned procedure. Another very important generator is the $^{82}$Sr/$^{82}$Rb pair, the strontium has a 25-day half-life, the rubidium daughter about 6 hours. This is one of the most useful isotopes for myocardial imaging (rubidium is a potassium analog), and is currently in very short supply. The $^{82}$Sr parent requires high-energy (~>70 MeV) proton beams, currently produced a few months of the year at large National Labs (Brookhaven and Los Alamos) or from accelerators in South Africa or Canada, that run their accelerators parasitically for isotope production. However, the total supply is not adequate for the growing demand for these generators.

Many valuable isotopes cannot be produced close to the clinical use-site. For these, rapid and efficient delivery systems are needed to move the isotopes from production to customer. Typical lifetime limits for this category of isotopes are of the order of a day or two. With air shipment channels, loss of activity is not unacceptable.

## Isotope Production

Neutron-rich radioisotopes are produced in reactors, by irradiation of target material with neutrons, or processing of spent fuel. Very few clinical sites will be fortunate enough to be close enough to a production reactor to not have to rely on an efficient transportation system. The most notable example is the previously-mentioned $^{99}$Mo/$^{99m}$Tc generator, which requires highly-enriched uranium feedstock, and is currently only available from three aging reactors, one in Canada, two in Europe. There is considerable concern in the nuclear medicine community about a secure source for this important isotope, and numerous proposals have been submitted to develop alternate sources for this isotope; involving either more complex processing required from non HEU cores, or by accelerator-driven systems [5,6]. This will be a long-range, and very expensive program.

Neutron-deficient isotopes are always produced with accelerators. Proton beams are used today, but deuteron and alpha beams could also play a role in the future. The nuclear reaction process most applicable is compound nucleus formation. The projectile is absorbed by the target nucleus (assuming it has enough energy in the center-of-mass of the target-projectile system to overcome the Coulomb barrier of about 5 MeV), in which the target nucleus is modeled as a liquid drop. The projectile brings its kinetic energy into the compound system, heating the liquid drop. The drop will lose its excess energy by various mechanisms, most usually by boiling off neutrons. This type of reaction is normally described as:

{Target} (p, xn) {Product}

Typically, each neutron will take off about 5-8 MeV of this excitation energy, so the neutron number of the desired reaction product will specify the beam energy required for most efficient production.

Note, protons can be emitted as well from the compound nucleus, however these must overcome the same Coulomb barrier to get out of the nucleus, so are less likely to be boiled off.[*]

Some examples: producing $^{18}$F, one starts with $^{18}$O as the target (in form of oxygen-18 water), producing the $^{18}$F via

$^{18}$O (p, n) $^{18}$F

Only one neutron is boiled off, so a low-energy beam is adequate. Beam energy must be higher than the Coulomb barrier, and must also add the higher Q value for the ground state of $^{18}$F (about 4.4 MeV), so a proton beam energy of around 10 MeV would be sufficient. One adds a few MeV to allow for beam windows, a thicker target and energy loss in the target to enhance production. Result is that $^{18}$F production is efficiently done with energies of 12-14 MeV.

---

[*] Cross sections for nuclear reactions involving proton emission can be quite high, comparable to neutron emission in some cases, but the mechanism is a "direct reaction" in which the incident proton collides with a proton in the nucleus and knocks it out, the projectile usually exiting as well if its energy is high. This would be a (p, 2p) reaction.

Another example is the above-mentioned $^{82}$Sr. The feedstock (target) material is natural $^{85}$Rb (100% natural abundance of this isotope… more on this later), but to get 82 nucleons we must boil off 4 neutrons.

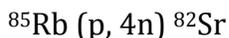
$$^{85}\text{Rb (p, 4n) }^{82}\text{Sr}$$

The lowest energy at which this reaction will occur is about 40 MeV, and again to maximize production efficiency a thick target is used, so the conventional wisdom is that one needs incident proton beams of about 60 - 70 MeV to efficiently produce this isotope.

**Cyclotrons for Isotope Production:**

As one increases the energy of the beam, more isotopes become accessible, however then economic and technology factors begin playing an important role. In today's market, two beam energy ranges have become common in the commercial isotope field: 12-15 MeV and 30 MeV. The first can be accessed by small cyclotrons, compact enough to fit in the basement of a large clinic to produce PET isotopes as needed. Units are shielded with a concrete clam-shell, entire assembly is about 3 meters across (outside shield dimensions) and 2 meters high. Power requirements are minimal, controls are all automatic, including the conversion of the isotope into the pharmaceutical used for the procedures. Costs for the entire system are no more than a few million dollars.

30 MeV cyclotrons from IBA (Belgium), TRIUMF-ACS (Canada), and a few other companies are the workhorses of isotope production "factories." One finds these machines located at centers that have several cyclotrons and the hot cells for processing the target material into pharmaceuticals. The centers will be located close to transportation hubs so that air delivery of isotopes can take place efficiently. These machines, with modest radiochemical cells, cost around $15M.

Higher energies, for currently difficult-to-obtain isotopes (such as the $^{82}$Sr/$^{82}$Rb generator) currently come from the National Labs, as stated before, or from parasitic operations at TRIUMF (Canada) and iThemba Labs (South Africa). However, no machines dedicated to the commercial production of this isotope are in operation today. While the Sr/Rb has received all the regulatory approvals for clinical use today, many new isotopes that could be produced with higher-energy beams are not yet in the production pipeline.

Developing commercial capability is a complicated process. Companies will invest in the technology if there is a market. FDA approval for clinical use is a *sine qua non* for enabling a market, and this approval comes only after extensive research. For the research to take place, the isotope must be available 12 months of the year, but the revenue from these isotopes for the research projects is not adequate to fund the development of the technology and the acquisition of the necessary accelerators. The strategy so far has been to rely on the physics laboratory accelerators that produce limited quantities of these research isotopes to establish the basis for FDA approval that would then open the market gates. This is proceeding, but at a slow rate.

Two companies have entered the 70 MeV cyclotron area: IBA has installed its C-70 cyclotron at ARRONAX in Nantes, France [7], and Best Cyclotrons has committed to deliver a 70-MeV cyclotron to the INFN Laboratory at Legnaro, Italy [8]. These are first steps into this field. Two National Labs have expressed interest in developing or acquiring cyclotrons in this energy range to be dedicated to radioisotope production, Brookhaven [9] and Oak Ridge [10], but have yet to receive funding for these initiatives. The technology for compact cyclotrons in this energy range is still in its infancy, not only for the cyclotron itself, but also for targetry, which will be discussed next.

**Targets**:

The amount of isotope produced depends on the amount of target material exposed to the beam, the cross section for producing the desired isotope, and the amount of beam current. The factors determining the cross section were discussed above. Typically one determines what is called an "excitation function" for producing the isotope, namely the range from the lowest energy where the desired nuclear reaction will occur at a reasonable rate (maybe one tenth of the peak cross section), to the highest energy on the other side of the peak. This excitation function will typically be about 10 to 20 MeV wide, so for instance the $^{82}$Sr product might have good production efficiency for protons between 50 and 70 MeV. The best target thickness then would allow for 20 MeV energy loss of the protons through normal dE/dx mechanisms (collisions with electrons). This could be a few mm of thickness.

Target material comes into play. To get a pure reaction product it will be usually best to start with a mono-isotopic target. Studying the Chart of Nuclides, one can quickly see that odd-Z nuclei have very few stable isotopes, usually one or two at most. Even-Z isotopes, on the other hand, have many. Tin, for instance, has 10 stable isotopes, while neighboring indium and antimony have 2 each. Good isotope production will usually require a separated-isotope target, which for even-Z elements could be quite expensive, particularly if one wanted a target of several mm thickness.

Developing good alpha beams could play a role in this, by allowing the option of bringing two protons into the compound nucleus instead of just one, one has greater flexibility in selecting target materials, with possibilities of using odd-Z feedstock instead of expensive separated even-Z isotopes that might be needed with a proton beam for producing an odd-Z isotope.

Another factor is heat dissipation. A 1 mA beam at 30 MeV carries 30 kW of beam power. This is a non-trivial heat load, which along with the radioactivity and neutron production present formidable challenges to the engineering design of good targets. This is, in fact, about at the frontier of isotope targets today. Cyclotrons on the market today produce about this level of current (within about a factor of 2 up or down), and present targets can just handle this heat load.

One technique for utilizing more beam current is to have multiple extraction points from the cyclotron, and deliver beam to two target stations simultaneously, sharing the beam power available from a high-current cyclotron. This is possible with an H⁻ accelerator equipped with two stripping foil extraction systems.

However, if target development could allow for higher heat dissipation, production of isotopes could be substantially increased. Specialized targets absorbing up to megawatts have been developed for various applications (e.g. spallation neutron sources, neutrino production, ADS applications, fusion reactors), so given the proper engineering it is not outside the realm of possibility that isotope targets could also be designed to handle substantially higher heat loads.

## Relationship with IsoDAR/DAEδALUS:

The DAEδALUS collaboration aims to address timely and important questions in neutrino physics, by developing compact and relatively inexpensive sources of neutrinos that can be located close to large (hundred-kiloton-class) hydrogen-bearing neutrino detectors [11]. The enabling technology being pursued is high-current cyclotrons, providing currents and beam-powers far in excess of present-day capabilities [12,13]. Key to this technology breakthrough is accelerating $H_2^+$ ions instead of protons; using this ion alleviates space-charge limitations at the injection point of the first cyclotron, and allows for stripping extraction at the high-energy (800 MeV) end for meeting the extremely low beam-loss requirement (specified at less than 1 part in $10^4$) for the very high power beams being produced. Anticipated is 10 milliamps of proton beams on target, for a peak power of 8 MW at full energy.

Each accelerator module for the DAEδALUS experiment (3 are planned) will contain an injector cyclotron (the DIC or DAEδALUS Injector Cyclotron); a normal-conducting 60 MeV/amu monolithic machine of about 4 meters' diameter, with a conventional electrostatic septum extraction system; followed by a superconducting separated-sector cyclotron (the DSRC or DAEδALUS Superconducting Ring Cyclotron) accelerating $H_2^+$ ions to 800 MeV/amu. Stripping foils dissociate the $H_2^+$ into two protons, which spiral inwards, and because of the very non-homogeneous magnetic field map can be made to cleanly exit the machine for transport into the target.

The DIC as a stand-alone cyclotron is also being viewed as an excellent choice for an experiment called IsoDAR (Isotope Decay At Rest), where the 60 MeV proton beam will be used to produce $^8$Li, a good source of electron antineutrinos for conducting a high-efficiency sterile neutrino search [14]. For this experiment, the additional possibility exists, should conventional electrostatic septum extraction prove difficult, of stripping extraction of the beam, as the $H_2^+$ molecule is not required for further acceleration. The 10 mA proton beam extracted deposits 600 kW on the production target, and can operate at 100% duty factor, yielding a highly attractive antineutrino source expected to provide discovery science in a few months of running.

The DIC developed for IsoDAR can clearly serve as a highly efficient isotope-producing machine for the 60-70 MeV niche discussed above. While the DIC design at 60 MeV/amu could be used directly, the design concepts could without much effort be adapted to a slightly higher energy, at 70 MeV/amu or even higher, should this be more desirable for isotope production.

Even if isotope-production targets are not available to take the full 600 kW of beam power, the extracted beam of $H_2^+$ can be manipulated to allow several stripping stations along the beam line for splitting beam to a number of target stations simultaneously. This will allow target flexibility for producing larger quantities of the same isotope, or several different isotopes at the same time.

This symbiotic relationship between medical applications and neutrino physics is very attractive for providing a path forward for development of the technology, and it is expected that as a result industrial partners will not be difficult to find in assisting with the development of the accelerators and targets required for the neutrino experiments.

## References

1. "Accelerators for Medicine"/"Radioisotopes", Chapter 2 section 1 in "Accelerators for America's Future", Workshop report, W. Henning, C. Shank convenors, Washington DC, October 2009. Report downloadable at http://www.acceleratorsamerica.org [link checked 8/30/2012]

2. "Workshop on the Nation's Needs for Isotopes: Present and Future", Rockville, MD, Aug 5-7 2008, J. Norenberg, P. Staples, R. Atcher, R. Tribble, J. Faught, L. Riedinger eds. US DOE/SC-0107.

3. "Compelling Research Opportunities using Isotopes, Final Report; One of Two 2008 Charges to NSCA on the National Isotopes Production and Applications Program", NSAC Isotopes Subcommittee, April 23 2009. Report available from http://science.energy.gov/np/nsac/reports/ (link checked 8/30/2012).

4. "Isotopes for the Nation's Future: A Long Range Plan: Final Report; Second of Two 2008 NSAC Charges on the Isotope Development and Production for Research and Applications Program" NSAC Isotopes Subcommittee, August 27, 2009. Report available from http://science.energy.gov/np/nsac/reports/ (link checked 8/30/2012).

5. "Accelerator-driven Production of Medical Isotopes", Workshop hosted by Cockcroft Institute, Science & Technology Facilities Council, UK, Dec 8-9, 2011. Talks available at, https://eventbooking.stfc.ac.uk/news-events/accelerator-driven-production-of-medical-isotopes?agenda=1 [link accessed 9/2/2012]

6. "Making Medical Isotopes: Report of the Task Force on Alternatives for Medical-Isotope Production." A. Fong, T. Meyer Eds, TRIUMF report (2008)

7. J. Martino for the ARRONAX Cyclotron Group, "ARRONAX*, a High Intensity Cyclotron in Nantes", Cyclotrons and their Applications 2007, http://accelconf.web.cern.ch/AccelConf/c07/PAPERS/215.pdf. [link checked 9/1/2012] (Medical isotope laboratory based on IBA C-70 cyclotron)

8  G. Moschini et al, "A Cyclotron Isotope Production Center for Biomedical Research" INFN-LNL-225 (2008), http://pcbat1.mi.infn.it/~battist/infn-med/T**Cyclotron**_LNL-XP.pdf. [link checked 9/1/2012] (Proposal for 70 MeV cyclotron for INFN-Legnaro, contract awarded to Best Cyclotrons 2011)

9. "Cost/Benefit Comparison for 45 MeV and 70 MeV Cyclotrons" May 26, 2005, study conducted by Jupiter Corp. for US DOE (Brookhaven National Lab). Report available from http://www.isotopes.gov/outreach/reports/Cyclotron.pdf (Link checked 8/30/2012).

10. "Holifield Radioactive Ion Beam Facility Cyclotron Driver White Paper", Oak Ridge National Laboratory Physics Division report, Nov. 3, 2008. Report available from: http://www.phy.ornl.gov/hribf/initiatives/cyclotron/cyclotron-upgrade.pdf (link checked 8/30/2012).

11. J.R. Alonso for the DAEδALUS Collaboration, "High Power, High Energy Cyclotrons for Decay-At-Rest Neutrino Sources: The DAEδALUS Project", arXiv 1109.6861 (2011).

12. L. Calabretta et al, "Preliminary Design Study of High-Power $H_2^+$ Cyclotrons for the DAEδALUS Experiment", arXiv 1107.0652 (2012).

13. M. Abs et al, "Multimegawatt DAEδALUS Cyclotrons for Neutrino Physics", arXiv 1207:4895 (2012), submitted to Nucl. Instrum. Meth. A.

14. A. Bungau et al, "An Electron Antineutrino Disappearance Search Using High-Rate $^8$Li Production and Decay", arXiv 1205:4419 (2012).